# High-efficiency broadband active metasurfaces via reversible metal electrodeposition

Qizhang Li[1], Sachin Prashant Kulkarni[2], Chenxi Sui[1], Ting-Hsuan Chen[1,3], Gangbin Yan[1], Ronghui Wu[1], Wen Chen[1], Pei-Jan Hung[1], Xubing Wu[1], Tadej Emersic[1], Koray Aydin[2], Po-Chun Hsu[1*]

[1]Pritzker School of Molecular Engineering, University of Chicago, Chicago, IL 60637, USA

[2]Department of Electrical and Computer Engineering, Northwestern University, Evanston, IL 60208, USA

[3]Thomas Lord Department of Mechanical Engineering and Materials Science, Duke University, Durham, NC 27708, USA

Email: pochunhsu@uchicago.edu

**Teaser:** Reversible metal electrodeposition enables active metasurfaces with exceptional tunability, spanning from visible to mid-infrared.

## ABSTRACT

Realizing active metasurfaces with substantial tunability is important for many applications but remains challenging due to difficulties in dynamically tuning light-matter interactions at subwavelength scales. Here, we introduce reversible metal electrodeposition as a versatile approach for enabling active metasurfaces with exceptional tunability across a broad bandwidth. As a proof of concept, we demonstrate a dynamic beam-steering device by performing reversible copper (Cu) electrodeposition on a reflective gradient metasurface composed of metal-insulator-metal resonators. By applying different voltages, the Cu atoms can be uniformly and reversibly electrodeposited and stripped around the resonators, effectively controlling the gap-surface plasmon resonances and steering the reflected light. This process experimentally achieved >90% diffraction efficiencies and >60% reflection efficiencies in both specular and anomalous modes, even after thousands of cycles. Moreover, these high efficiencies can be extended from the visible to the near- and mid-infrared regimes, demonstrating the broad versatility of this approach in enabling various active optical and thermal devices with different working wavelengths and bandwidths.

**INTRODUCTION**

Tailoring light-matter interactions at subwavelength scales has garnered tremendous scientific interest and is key to a plethora of transformative technologies such as invisible cloaks (*1–3*), flat lenses (*4–6*), advanced waveguides (*7, 8*), and daytime radiative coolers (*9–15*). Its great potential has recently been manifested in metasurfaces which provide a full manipulation of light on the amplitude, phase, and polarization within a subwavelength-structured interface (*16–20*). With the promise to achieve the ultimate miniaturization of optical elements, metasurfaces have rapidly advanced to the forefront of optics research over the past decade (*21–23*). Furthermore, the emergence of metasurfaces also raises the hope of revolutionizing active optical components for advanced applications that demand ultracompact and lightweight optics, such as augmented/virtual reality glasses and unmanned aerial vehicles (*24–43*). Nevertheless, it is still challenging to achieve active metasurfaces with substantial tunability compared to conventional bulky optics like deformable mirrors (*44*). Addressing this challenge is essential for active metasurfaces to outperform their bulky counterparts and step further toward commercial success (*45, 46*).

An essential pathway to achieve high-contrast active metasurfaces is to use materials with highly tunable optical properties. Among the most promising candidates are phase change materials which feature the metal-insulator transition (MIT), capable of creating substantial changes in refractive index (*47*). To date, MIT materials such as vanadium dioxide (*14*, *48–52*) and conjugated polymers (*31*, *53–55*) have demonstrated many successful applications in active optical and thermal devices. However, MIT materials exhibit much lower plasma frequencies in their metallic state compared to metals (e.g., Cu), therefore limiting their optical tunability, especially at short wavelengths (*56*). The ideal MIT scenario with the highest accessible optical contrast would be creating metals from vacuum and making them vanish as desired. Such a scenario is not physically feasible due to the law of mass conservation. Nevertheless, this effect can be mimicked with reversible metal electrodeposition (RME), which allows for the creation and removal of metals on demand, leading to an effective switch between metals and electrolytes (*56*, *57*). In this way, RME achieves gigantic optical contrast across an ultrabroad wavelength range and has enabled a variety of applications, including electrochromic windows (*58*, *59*), visible and thermal camouflage (*60*, *61*), and tunable radiative cooling systems (*62*, *63*). Though appealing, these demonstrations were mostly done on thin film configurations, and the potential of RME in controlling light-matter interactions with subwavelength structures remains largely underdeveloped (*60*, *64*). Given that RME yields almost ideal optical contrast, it is of both fundamental and practical interest to explore the feasibility of RME within micro and nanostructures and its ability to control subwavelength light-matter interactions.

In this work, we experimentally explored the capability of RME performed on subwavelength structures for realizing high-contrast active metasurfaces. This concept is showcased by an active beam-steering metasurface with tunable metal-insulator-metal (MIM) resonators controlled by the RME of Cu. The top metal and the middle dielectric layers are structured to form the resonant antennas, whereas the bottom metal serves both as an optical element that reflects the incoming light, and as an electrical component utilized as a highly conductive working electrode to enable high-performance RME. As a result, the Cu atoms can be uniformly and reversibly deposited and stripped around the subwavelength resonators by switching the applied bias. This effectively controls the gap-surface plasmon resonance, leading to high-efficiency dynamic beam steering. Furthermore, such substantial tunability can be achieved across a very broad bandwidth from visible to mid-infrared (IR) ranges. Our results highlight promising opportunities to realize the high-contrast active metasurfaces demanded for emerging optical and thermal applications.

**RESULTS**

**Potential of RME for achieving substantial broadband tunability**

Figure 1A illustrates the working principle of leveraging RME to control the subwavelength light-matter interactions of a MIM resonator, which is immersed in an electrolyte with the working and counter electrodes. By applying a negative bias to the working electrode, a layer of high-reflectivity metal will be deposited around the resonator, which prevents incident light from entering the resonant cavity and turns off its plasmon resonance. The resonance can be fully recovered by reversing the bias polarity, which in turn dissolves the deposited metal into the electrolyte as metal ions. As shown in Fig. 1B, the effective control of gap-surface plasmon resonances extends from the visible to mid-IR ranges using resonators of different sizes. Such a substantial and broadband tunability of RME can be expected from the gigantic refractive index differences between the metals and the corresponding electrolytes over a broad bandwidth (Table S1). To further demonstrate the optical tunability potential of RME, we compare it with other representative active materials by calculating their tuning ranges in terms of the single-interface normal-incidence reflectivity, $R = |(\tilde{n} - 1)/(\tilde{n} + 1)|^2$ where $\tilde{n}$ is the complex refractive index (inset of Fig. 1C). This is a reasonable standard for comparison as it involves only the material properties (i.e., $\tilde{n}$). The ideal active material should feature $\Delta R = 1$, fully suppressing reflection at the interface ($R = 0$) in the initial state and entirely reflecting all the incident light ($R = 1$) as the material property changes. As shown in Fig. 1C, RME stands out from commonly used active materials with its near-unity contrast ($\Delta R \approx 1$) across the entire wavelength range of interest. This promises a great potential of RME to achieve active metasurfaces with substantial tunability over an ultrabroad wavelength range.

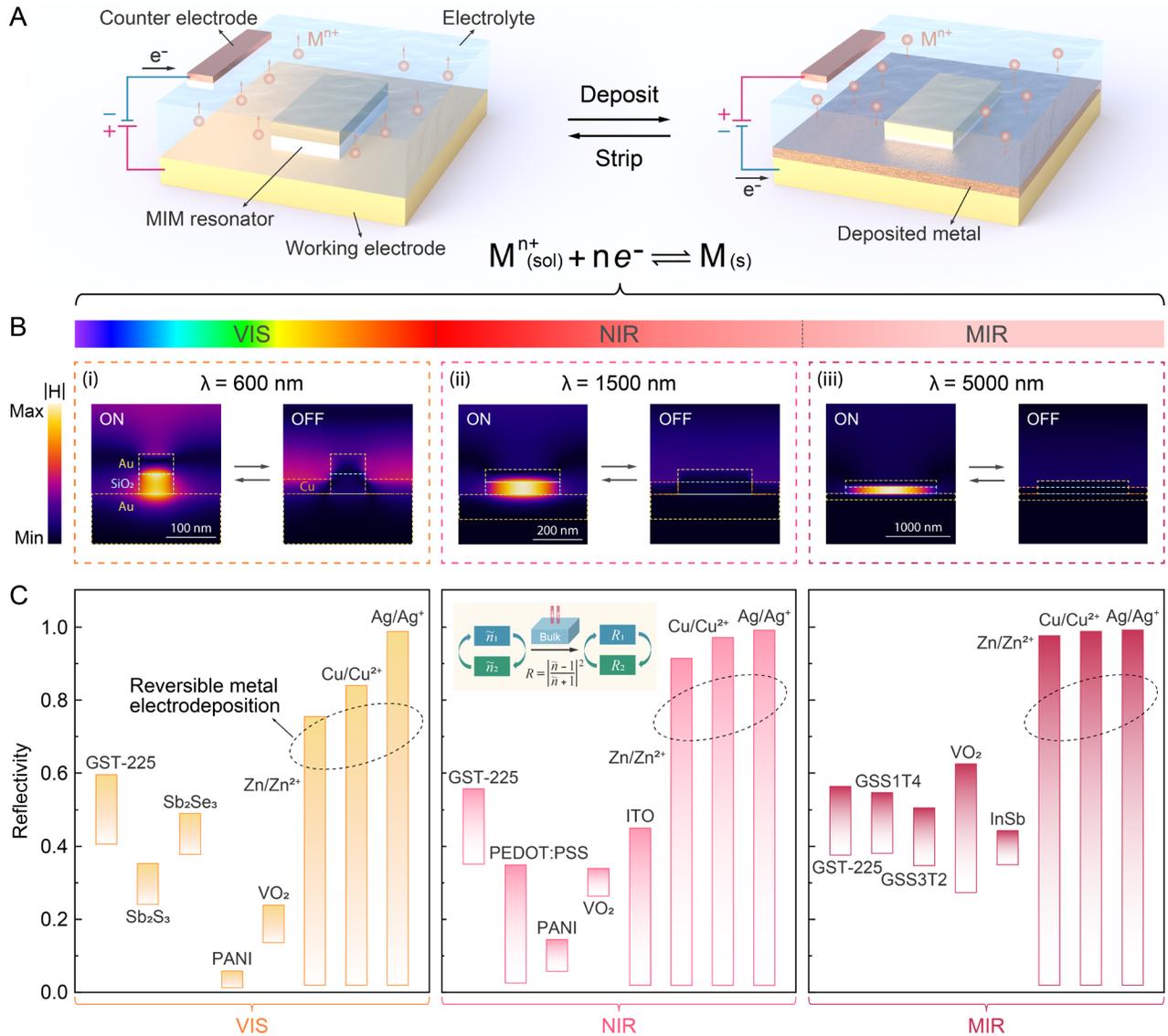

**Figure 1. Concept and potential of active metasurface with substantial broadband tunability enabled by RME.** (A) Schematic of a tunable MIM resonator for dynamic light manipulation via RME. (B) Simulated magnetic field profiles of the MIM resonators controlled by RME at the visible (i), near-IR (ii), and mid-IR (iii) regimes. Simulation details can be found in Fig. S1. (C) Calculated tunable range of single-interface reflectivity for RME and various active materials at different wavelengths, with RME demonstrating the highest tuning potential. The material optical properties used in calculations are provided in Table S1(*24*, *31*, *65–75*).

**High-efficiency dynamic beam steering via RME-based active metasurface**

To demonstrate the potential of RME-based active metasurfaces, we focus on dynamic beam steering, as achieving high-efficiency beam-steering metasurfaces has been quite challenging due to the limited optical property contrast of active materials (Table S2). As illustrated in Fig. 2A, our metasurface consists of MIM resonator arrays with a trapezoidal shape designed to support broadband resonances (*76, 77*). In its pristine state when there is no Cu deposition, these trapezoid-shaped antennas impart a phase gradient to the reflected light along the device surface, causing the reflection angle to deviate from the incident angle and resulting in anomalous reflection (*78, 79*). The anomalous reflection (corresponding to $1^{st}$ order diffraction) can be greatly suppressed by the metal electrodeposition, which eliminates the phase gradient and switches the reflected light to the specular direction ($0^{th}$ order). Therefore, an electrically tunable metasurface capable of steering light into different directions is realized by changing the applied bias to reversibly deposit and strip the metal layer.

For experimental demonstration, we specifically designed and fabricated the metasurface to operate at visible wavelengths. With minimum feature sizes on the order of tens of nanometers, a successful demonstration of this active metasurface would validate the practicality of RME in controlling light-matter interactions at nanoscales. The top view of the fabricated metasurface is shown in Fig. 2B by using scanning electron microscopy (SEM). We used gold (Au) and silicon dioxide ($SiO_2$) for the metal and insulator components of the MIM resonators, considering their robustness in nanofabrication and RME processes. For the RME operations, we chose Cu as the working metal for its great stability, long durability, and fast switching speed (*58, 63, 80*). By applying a negative voltage to the bottom metal of the metasurface with respect to the counter electrode of Cu foil, a Cu layer was densely and uniformly deposited on the Au substrate as characterized with SEM and energy-dispersive X-ray spectroscopy (EDS) in Fig. 2C. This was further confirmed with cross-sectional scanning transmission electron microscopy (STEM) images as shown in Figs. 2D and 2E. Such high-quality electrodeposition was enabled by both the high conductivity of the Au electrode and its energy-favorable interface to facilitate the nucleation of Cu (*61, 63, 81*). As indicated by the effective medium theory (*82*), the dense and continuous deposition morphology is essential to reach the percolation threshold for behaving as a metal film with the desired optical reflectivity (*63*). Consequently, the gap-surface plasmon resonance is turned off by the Cu deposition and the metasurface no longer supports anomalous reflection but rather behaves as a reflective mirror with only specular reflection (Fig. 2F). The metasurface fully regains its functionality when applying a positive voltage that quickly oxidizes Cu(s) to $Cu^{2+}$ cations dissolved in the electrolyte (Fig. 2F). Such an on/off switch can be done at a timescale of 500 ms when using an electrolyte with 1 M $Cu^{2+}$, as illustrated in Fig. 2G. More importantly, with the effective deposition and stripping of Cu at nanoscales, our active metasurface experimentally demonstrated ultrahigh diffraction efficiencies of >90% for both the anomalous and reflection modes. Here, the diffraction efficiency, $\eta_d$, is defined as the ratio of the light intensity in the desired diffraction order to the total reflected light intensity. This total intensity can be approximated by the sum of all three major diffraction orders ($0^{th}$ and $\pm 1^{st}$) as the contributions from higher orders are negligible (Fig. S2). It is noteworthy that achieving high diffraction efficiency is particularly important for metasurfaces to be utilized in practical applications that demand high signal-to-noise ratios (SNR) (*83–85*).

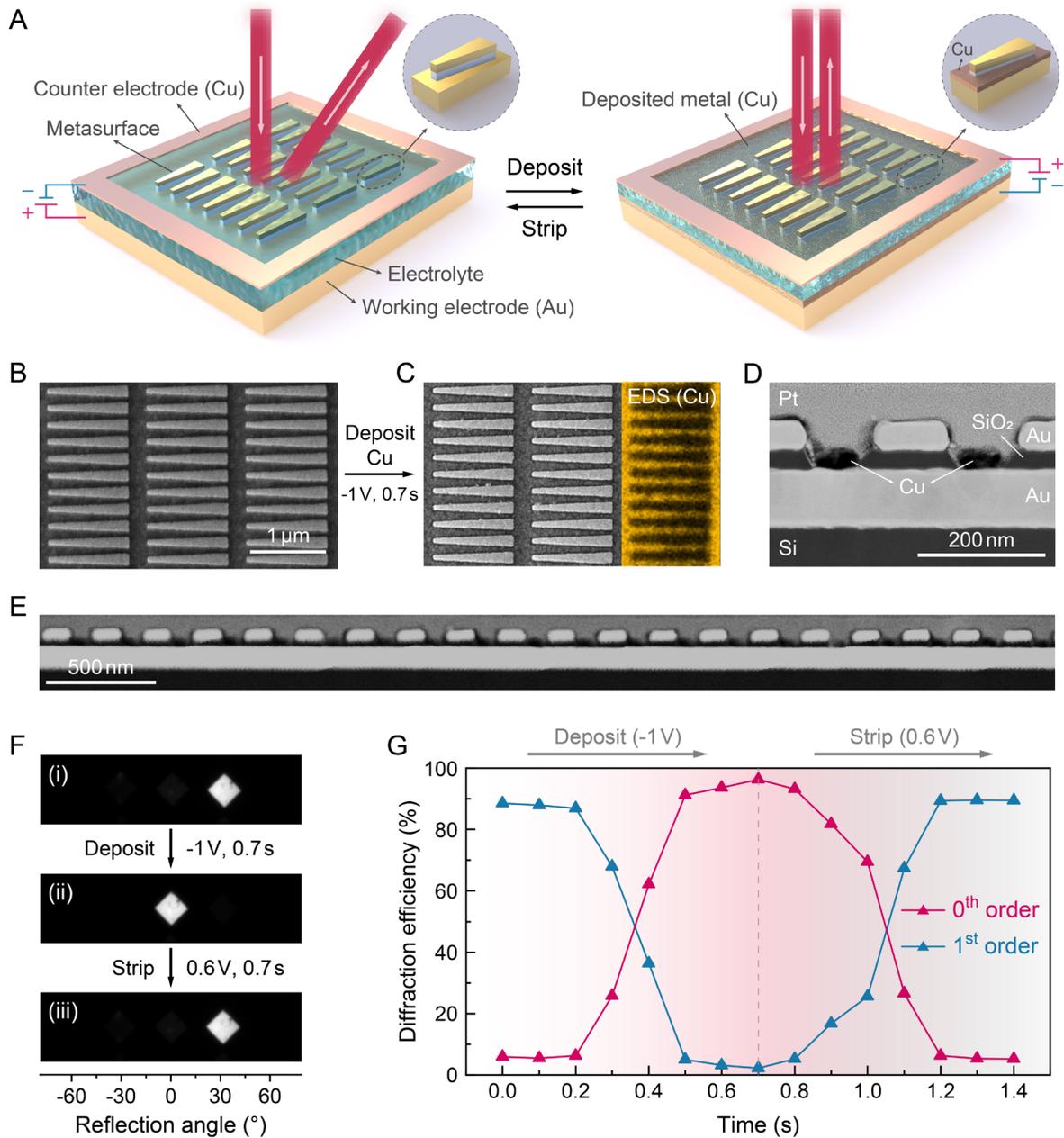

**Figure 2. Active beam-steering metasurface via reversible Cu electrodeposition.** (A) Schematic of the electrochemical device for dynamic beam steering. The device can be electrically switched between the anomalous reflection mode (Cu-stripped, left) and the specular reflection mode (Cu-deposited, right). (B and C) Top-view SEM images of the metasurface before (B) and after (C) Cu electrodeposition. The deposited Cu area is partly highlighted via EDS. (D and E) Cross-sectional

STEM images of the metasurface deposited with Cu. The sample lamella was prepared using focused ion beam (FIB). (F) Microscope Fourier imaging (*86*) of the reflected beam profiles for the metasurface as fabricated (i), after electrodepositing (ii), and after stripping (iii) of Cu. (G) The $0^{th}$ and $1^{st}$ order diffraction efficiencies when depositing and stripping Cu with different times.

**Mechanisms of active beam-steering metasurface**

The underlying mechanisms of our active metasurface were theoretically studied with full-field electromagnetic simulations using the unit cell depicted in Fig. 3A. In its initial state (prior to Cu deposition), the MIM resonator supports a gap-surface plasmon resonance localized near the center of the antenna when illuminated with 685-nm light, as indicated by the enhanced magnetic field within the MIM cavity (Fig. 3B). With the dramatic phase shift imparted by the MIM resonator, the trapezoidal antenna enables a near-2π phase coverage required for the anomalous reflection response, owing to its gradually increasing width along the antenna which induces different phase shifts (Fig. 3C). This phase gradient imparts an additional wave vector to the normally incident light and leads to anomalously reflected wavefront as shown in Fig. 3D. In great contrast, after depositing only a 30 nm Cu layer around the cavity, the plasmon resonance is significantly suppressed (Fig. 3B), fully eliminating the phase gradient (Fig. 3C). As a result, the reflected light is steered from the anomalous angle to the normal direction (Fig. 3D). It is noteworthy that the reflectivity remains consistently high when switching the beam propagation directions, as indicated by the reflectivity plots in Fig. 3C. This is the foundation to realize dynamic beam steering with high efficiencies without loss of incident optical power.

To further analyze how the reflected light is steered between different directions during electrodeposition, we simulated the far-field intensity profiles for varying Cu thicknesses (Fig. 3E). The simulation results agree well with the experimental observations (Fig. 3E insets), where the same nominal thickness of Cu is deposited. Both theoretical and experimental results demonstrate that, with a 40 nm SiO$_2$ cavity, a 30 nm Cu layer is sufficient to fully switch the beam propagating direction. This low requirement for deposited Cu thickness is crucial for achieving active metasurfaces with excellent durability and fast switching speeds (*87*). In addition, the required deposition thickness can be reduced to 20 nm by optimizing the MIM resonator design with a thinner SiO$_2$ layer, or even to 10 nm by embedding the bilayer antenna into the Au substrate (Figs. S3 and S4), indicating great promise for further improving the device performance.

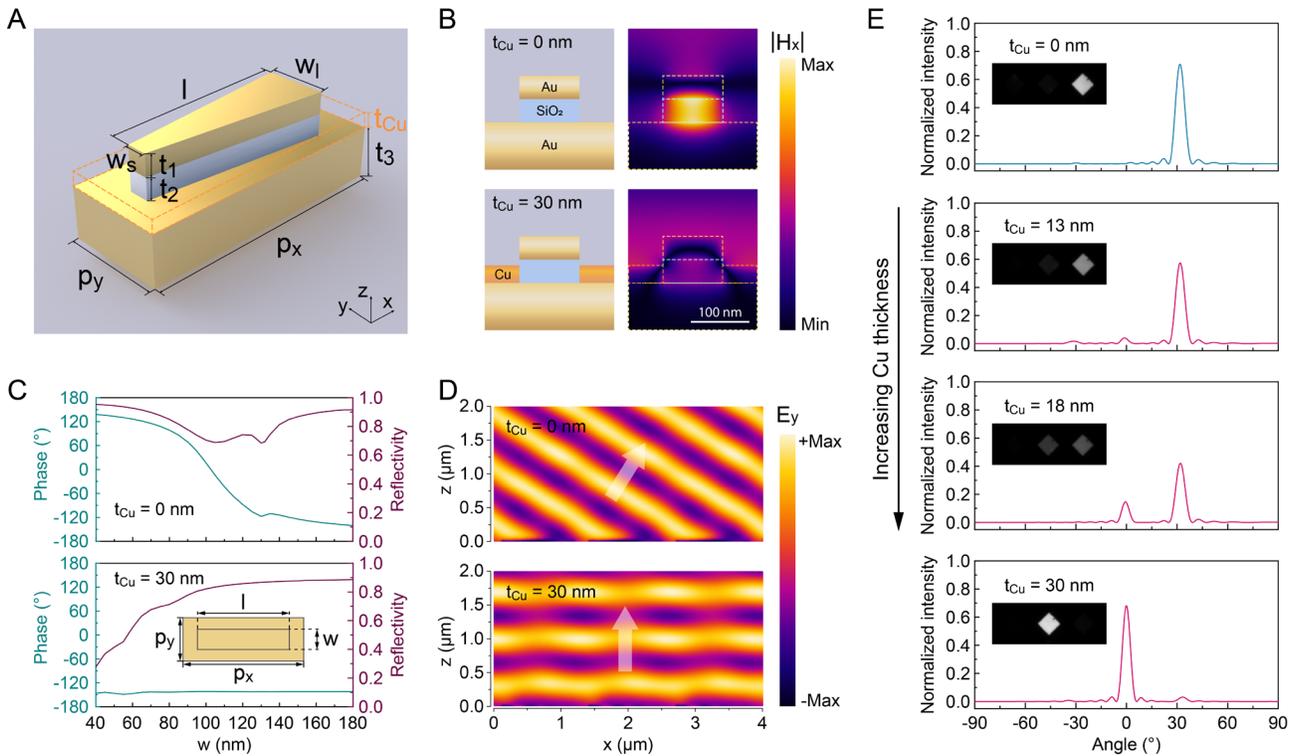

**Figure 3. Mechanisms of the switch between anomalous and specular reflection modes for the**

**RME-based active metasurface.** (A) Schematic of the unit cell with a trapezoidal antenna. For the design operating at visible wavelengths, the geometric parameters are $P_x = 1300\ nm$, $P_y = 220\ nm$, $l = 1056\ nm$, $w_l = 150\ nm$, $w_s = 55\ nm$, $t_1 = 40\ nm$, $t_2 = 40\ nm$, $t_3 = 100\ nm$. (B, C, and D) Comparisons between the metasurfaces with and without Cu electrodeposition in terms of the cross-sectional unit-cell schematics and corresponding simulated magnetic fields (B), phase and reflectivity profiles for a rectangular antenna as a function of width (C), and the electric fields of reflected light above the metasurface (D). The cross-section mapping is along the center of the unit cell ($P_x/2$). (E) Simulated far-field intensity profiles of reflected light with increasing Cu thickness, which is normalized to the peak reflection intensity of a silver mirror. Insets are corresponding measured beam profiles. In the simulations, the wavelength of light is 685 nm, and the polarization is along the *y*-axis.

**In-situ and ex-situ characterizations on beam steering performance**

To characterize the beam steering performance, we set up a microscopy system, as shown in Fig. 4A. Incident light with a wavelength of 685 nm was generated using a bandpass filter in the path of a white light source. An objective with a small numerical aperture (NA = 0.3) allowed us to clearly observe the distinct difference in the metasurface's appearance between the deposited and stripped states (Fig. 4B). When Cu was deposited on the device, the entire metasurface area (300×340 μm²) appeared as bright as the surrounding flat region. Conversely, when Cu was stripped from the device, the metasurface area became noticeably dark, as most of the light was anomalously reflected to ~30°, beyond the detectable angle range of the 0.3-NA objective (<17.5°).

The angle-resolved quantitative measurement of light intensity was further achieved by Fourier plane imaging (*86*), a technique that uses the Bertrand lens to project the Fourier transform of the image plane onto the camera, allowing each pixel to correspond to a specific reflection angle (Fig. S5). In this case, a high-NA objective was required to capture reflected light over a wide angular range. For in-situ measurements, we used a 0.8-NA dipping lens immersed in a dilute electrolyte containing 50 mM $Cu^{2+}$. While using dilute electrolytes decreases the switching speed, it ensures sufficient reflected light intensity for camera detection, compensating for the strong absorption of red light in the $Cu^{2+}$ electrolyte due to the long working distance of the dipping lens (3.3 mm). Figure 4C displays the line profile of the reflected beam intensity for the initial five cycles of the experiments, during which the light was efficiently steered toward different directions. By analyzing the light intensity at different beam spots, we derived the diffraction efficiencies for the $0^{th}$ and $1^{st}$-order diffractions in both the deposited and stripped states. As shown in Fig. 4D, both states stably achieved over 90% diffraction efficiencies for up to 600 cycles.

To further explore the reversibility of the device, we conducted a cycle test via ex-situ measurements on both diffraction and reflection efficiencies at different states. The reflection efficiency, $\eta_r$, is defined as the intensity of the targeted diffracted light normalized to the incident light intensity, which is approximated by the reflection from a silver mirror. As shown in Fig. 4E, the RME-based active metasurface consistently achieved ~60% reflection efficiency and ~90% diffraction efficiency without degradation even after 3000 cycles. This excellent cyclability is largely attributed to the use of the Au back reflector as the working electrode, which is ideal for RME due to its high conductivity and low lattice mismatch with Cu (*88*). Our findings highlight the potential of RME to enable active metasurfaces with exceptional optical performance, high stability, and long-term durability.

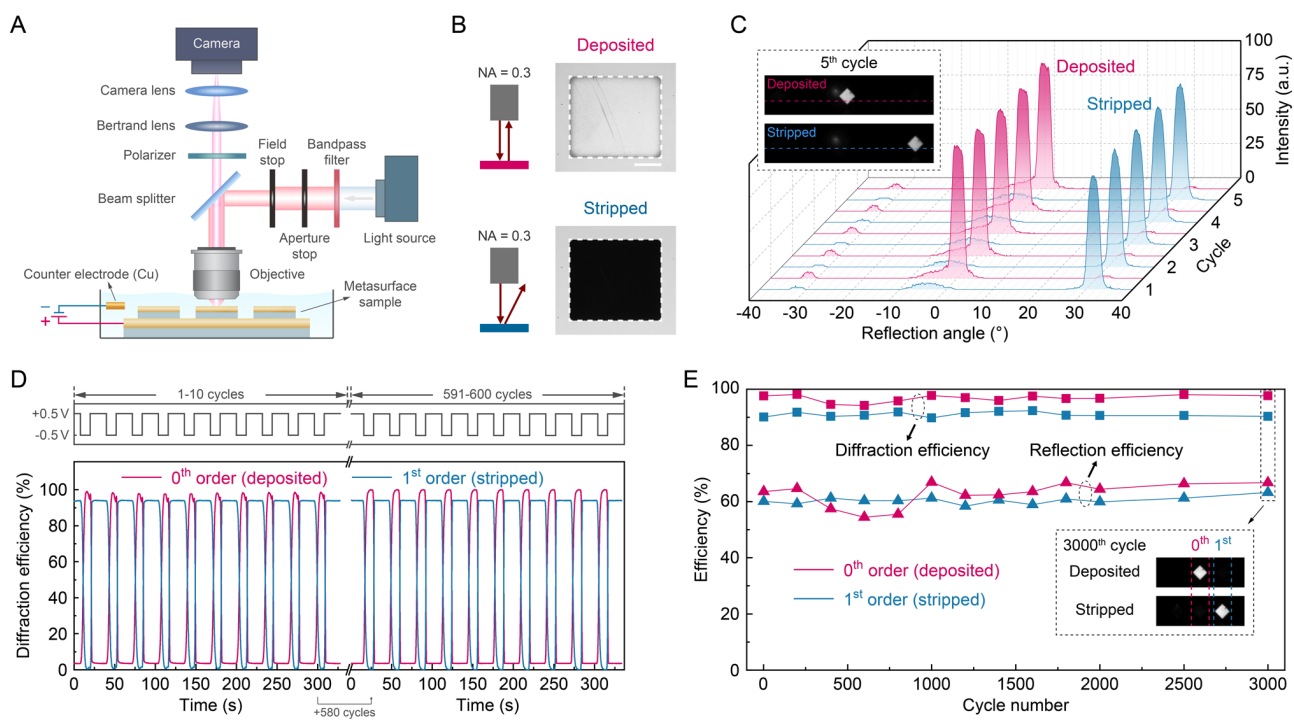

**Figure 4. Experimental demonstration of consistently high-efficiency dynamic beam steering.** (A) Schematic of the microscope setup for characterizing the reflected beam profile. A dipping lens is immersed in the working electrolyte for the in-situ measurements. (B) Microscope images of the metasurface sample in the deposited and stripped states when a 0.3-NA objective is used. The metasurface area is circled with a white dashed line. The scale bar is 100 μm. (C) Line profiles of intensity captured via the Fourier plane imaging for the initial five in-situ cycles. Insets are the camera images for the fifth cycle, and the dashed lines indicate where the intensity profile is extracted. (D) In-situ measurement of the $0^{th}$- and $1^{st}$-order diffraction efficiencies in the deposited and stripped states for the 1-10 (left) and 591-600 (right) cycles. The corresponding videos are provided as supplemental auxiliary files. (E) Cycle life test of the active metasurface on the diffraction and reflection efficiencies. The inset shows the Fourier plane images for the deposited and stripped states at the $3000^{th}$ cycle. The test was done in ex-situ experiments by applying -1 V for 0.7 s in the deposition process and 0.5 V for 10 s for a complete stripping.

**Broadband demonstration of RME-based active metasurfaces**

The synergy of Cu as a Drude metal and the trapezoidal antenna design suggests substantial broadband capabilities for our active beam-steering metasurfaces. To explore this potential in the visible and near-IR regimes, we employed a custom-built angle-resolved spectroscopy system to characterize the anomalously reflected light, as depicted in Fig. 5A (*76, 89*). Subsequently, the diffraction and reflection efficiencies for different diffraction orders were determined with a microscope-based spectroscopy system. For the visible regime, we continued using the aforementioned metasurface design and investigated its broadband performance at other visible wavelengths. As plotted in Figs. 5B and S6, extraordinary optical performance ($\eta_d \approx 90\%$ and $\eta_r \approx 60\%$) was observed for both deposited and stripped states across a 100-nm-wide wavelength range from 610 to 710 nm. This broadband tunability arises from the leverage of trapezoidal antennas (Fig. S7) and the substantial optical contrast provided by the REM of Cu (Fig. 1C). We note that Cu is less suitable for devices working at shorter visible wavelengths due to its absorptive nature in that region, and RME of silver or aluminum can be more effective (*61, 90*). For the near-IR demonstration, we redesigned and fabricated the metasurface with larger feature sizes to accommodate wavelengths up to 800 nm, considering the experimental setup availability. The measurement results, also shown in Fig. 5B, consistently demonstrated high efficiencies across this extended wavelength range.

For the mid-IR wavelengths, the angle-resolved reflectance spectrum was measured with an IR ellipsometer as shown in Fig. 5C. Due to the limitations of the instrument on detection angles, we designed the mid-IR metasurface considering a tilted incidence angle of 32° and conducted measurements at reflection angles of 32° and above. In this case, the $0^{th}$ and $1^{st}$ order diffracted light with a 3 μm wavelength were predicted to feature reflection angles of 32° and 50°, respectively, according to the generalized Snell's law (*78*), which was confirmed by the measurements (Fig. S8). The reflection efficiencies of the $0^{th}$ and $1^{st}$-order diffractions for both deposited and stripped states are shown in Fig. 5D, where $\eta_r \approx 60\%$ is consistently achieved over a broad wavelength range from 2.6 to 3.6 μm. Although the diffraction efficiencies at these wavelengths could not be determined due to the difficulties in measuring the -$1^{st}$ order diffraction, one can still expect considerably high values from the distinct intensity contrast between $0^{th}$ and $1^{st}$ orders.

Using the measured beam-steering efficiencies, we determined the SNR of the RME-based metasurfaces, a key figure of merit for real-world applications that require the output signal to be distinguishable from the background (*34*). The SNR is the intensity ratio between the dominant diffraction order and the second-highest one (*34*). Since the active metasurfaces exhibit different SNRs during switching, the smallest value among different states is considered, as it represents the performance limit of active devices. Our results show that RME-based metasurfaces consistently exhibit higher SNR from visible to mid-IR ranges compared to many other active-material systems (Fig. 5E), in line with the optical contrast comparison in Fig. 1C. While the reflection efficiencies might be limited in the mid-IR regime due to the electrolyte absorption, the high diffraction efficiencies and SNR would remain unaffected. We anticipate the great potential of RME-based metasurfaces could intrigue further research efforts to demonstrate more possibilities such as faster switching speed (*24*, *41*, *91*), multiple-angle steering (*34*, *92–94*), high-performance transmissive counterparts (*26, 29, 31, 95, 96*), and novel functionalities (*33, 97–99*).

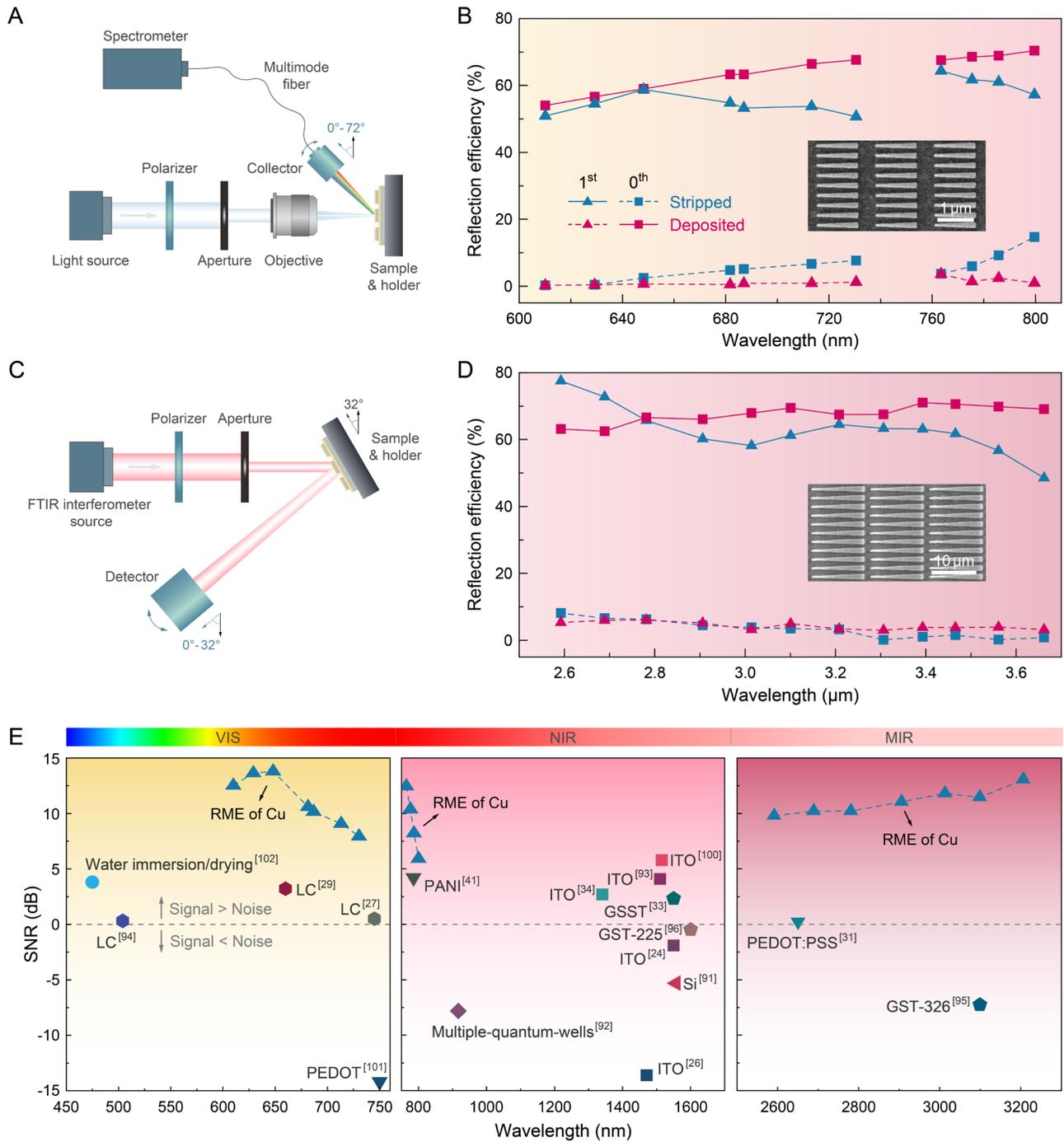

**Figure 5. RME-based metasurfaces with broadband dynamic beam-steering capability.** (A and B) Schematic of the angle-resolved spectroscopy system (A) and the ex-situ measured reflection efficiencies in the visible and near-IR ranges (B). Inset is the SEM image of the near-IR sample. (C

and D) Schematic of the ellipsometer setup (C) and the ex-situ measured reflection efficiencies in the mid-IR range (D). Inset is the SEM image of the mid-IR sample. (E) Accessible SNR of the RME-based metasurfaces and other up-to-date beam-steering devices working at different wavelengths (*24, 26, 27, 29, 31, 33, 34, 41, 91–96, 100–102*).

**DISCUSSION**

We have demonstrated the remarkable potential of reversible metal electrodeposition for enabling active metasurfaces with exceptional broadband tunability. Leveraging the ideal optical contrast provided by RME, we experimentally realized active beam-steering metasurfaces with reflection efficiencies of >60% and diffraction efficiencies of >90%, leading to remarkable SNRs up to ~15 dB. Furthermore, such high performance well extends across the visible, near-IR, and mid-IR ranges using different-sized metasurfaces based on the same design principle. This was accomplished through a plasmonic metasurface design that combines both photonic and electrochemical considerations, utilizing the metal back reflector of the antennas as a desirable working electrode for RME operation. Consequently, Cu atoms can be uniformly and reversibly electrodeposited and stripped at the nanoscale during switching, leading to exceptional optical performance and electrochemical sustainability, as validated by the ex- and in-situ experiments. This work opens opportunities for dynamic control of light at subwavelength scales, facilitating the development of active optical and thermal metadevices with ultrahigh efficiencies.

**MATERIALS AND METHODS**

Numerical simulations

The numerical simulations of the metasurfaces are performed with the commercial software of COMSOL version 5.6. We apply periodic boundary conditions to the *x*- and *y*-directions at the surface edges of the 3D unit cell model, which is normally illuminated by an incident plane wave with a polarization along the *y*-direction. The perfectly matched layer is added to both the top and bottom ends of the unit cell for convergence purposes. The refractive index of Au and $SiO_2$ in simulations is obtained from the ellipsometry measurements, while that of Cu is adopted from the literature (*66*).

Fabrication of metasurfaces for operating in the visible and near-IR regimes

The metasurfaces with nanoscale feature sizes were fabricated using electron-beam lithography. We started with covering a glass substrate with a 5 nm chromium (Cr) adhesive layer and a 100 nm Au layer using electron-beam evaporation (AJA e-beam Evaporator). The sample was primed with HMDS for 5 min, spin-coated with a 950 PMMA A4 resist at a rate of 6000 rpm for 1 min, baked at 180 °C for 5 min, and then exposed in the Raith EBPG5200 E-Beam lithography system operated at 100 kV with a dose of 700 µC/cm². The sample was developed in a MIBK/IPA (1:3) solvent for 2 min at room temperature and then rinsed with IPA and DI water for 1 min each. The antenna pattern was formed by sequentially depositing $SiO_2$ and Au using electron-beam evaporation. Before and after the $SiO_2$ deposition, 2 nm of Cr was deposited to enhance the antenna adhesion. Note that a 3 nm layer of $SiO_2$ was deposited on top of the Au antenna to prevent the undesired Cu electrodeposition. Afterwards, the sample was lifted off by soaking in acetone at 60 °C for 15 min and then sonicating for 10 s. The sample was then subjected to $O_2$ plasma processing to completely remove the residual resist.

Fabrication of metasurfaces for operating in the mid-IR regimes

The micro-size metasurfaces were fabricated by a typical photolithography process. A silicon wafer substrate was covered with a 5 nm Cr layer and a 100 nm Au film using electron-beam evaporation. The subsrate was primed with HMDS for 5 min and then spin-coated with a negative photoresist (AZ nLOF 2020) at 4000 rpm for 45 s. Afterwards, the sample was baked at 110 °C for 1 min and then exposed to a 375 nm laser source in the Heidelberg MLA150 lithography system with a dose of 260 mJ/cm$^2$. The exposed sample was further baked at 110 °C for 1 min, developed in the AZ 300 MIF developer for 1 min, and rinsed with DI water for 1 min. The subsequent processes and instruments are the same as those used in the electron-beam lithography, including the deposition of antenna materials and the lift-off process.

Electrochemical setup

The electrolyte consisted of 1 M $Cu(ClO_4)_2$, 1 M $LiClO_4$, and 20 mM $HClO_4$. $Cu(ClO_4)_2$ was used because $ClO_4^-$ is the desirable anion for aqueous reversible Cu electrodeposition that suppresses the side reactions (i.e., forming insoluble Cu complexes) (*80*). $LiClO_4$ serves as the supporting electrolyte to increase solution conductivity, and the $HClO_4$ prevents the deposition of insoluble $Cu_2O$ complexes (*80*). The electrochemical experiments were conducted with the VMP3 potentiostat (BioLogic) under constant potential conditions. The experiments were carried out with the 2-electrode system where the metasurface served as the working electrode and the Cu foil was the counter electrode.

## Optical measurements with microscopy system

The microscopy system (Fig. 4A) was built upon the Olympus BX53 microscope which could provide the visualization of the dynamic beam steering towards different reflected angles. A white LED light source was used for illumination, filtered by a bandpass filter at 685 nm with a bandwidth of 10 nm (AVR Optics). The aperture stop in the Köhler illumination tube was used for better collimation of light, and the field stop for limiting the size of the illuminating area. We note that the diamond shape of the beam spot arose from the shape of the aperture stop that we used. The desired polarization of the light was achieved by a polarizer after the reflection light was collected with the objective. The Fourier plane imaging was realized by using the Bertrand lens, a standard lens placed in between the objective and the camera lens. In the ex-situ experiments, two objectives (Olympus, 10×, NA = 0.3; Olympus, 100×, NA = 0.9) were used for different imaging purposes as illustrated in the main text. The dipping objective (Leica, 40×, NA = 0.8) was used for the in-situ experiments.

## Broadband beam-steering characterization for visible and near-IR wavelengths

The intensity of the anomalously reflected beam at different visible and near-IR wavelengths was measured with a custom-built angle-resolved spectroscopy system (Fig. 5A). The light from a broadband halogen lamp source was collimated and polarized before entering the aperture to achieve a desired incident beam feature. The light was focused on the sample using the 4× Nikon achromatic objective and then collected at different reflection angles by another objective installed on a rotatable stage. With this setup, the accessible range of the reflection angles was from 18° to 90°. By using the multimode fiber, the collected light was then coupled to a spectrometer system, which consists of a 303 mm focal-length monochromator and Andor−Newton electron multiplication CCD.

To further characterize the reflection and diffraction efficiencies of different diffraction orders, we employed a microscope-based spectroscopy system (HORIBA SMS Microspectroscopy). This instrument utilizes a series of mirror arrays to collimate and shrink down the beam spot size of a tungsten halogen white light source (Ocean Optics, 4.7 mW, 360-2400 nm). The 0-order reflection light intensity was measured by using a 0.3-NA objective which can eliminate the higher-order reflections. By using a silver mirror to characterize the incident light intensity, we then derived the $0^{th}$-order reflection efficiency. The total reflection efficiency (approximated to the sum of $0^{th}$ and $\pm 1^{st}$ orders) was similarly characterized but with a 0.9-NA objective. In combination with the measurement results from the angle-resolved spectroscopy system, we could then distinguish the reflection efficiency of $0^{th}$, $+1^{st}$, and $-1^{st}$, respectively.

Broadband beam-steering characterization for mid-IR wavelengths

The mid-IR measurements were performed on the IR-VASE ellipsometer (J.A. Woollam), which includes a Fourier transform infrared (FTIR) spectroscopy interferometer light source and a deuterated triglycine sulfate (DTGS) detector installed on a rotatable stage (Fig. 5C). Due to the limitation of the detectable reflection angles, the measurements were conducted with a tilted incident angle of 32°, and the reflected light was collected starting from 32° to larger reflection angles. During the experiments, the iris of the light source was turned to the smallest value (~1 mm) to yield a small beam spot to be fully covered in the patterned area (13×14 mm$^2$). An evaporated Au mirror was used as a reference sample for calculating reflection efficiency, which was measured at 32° for both incident and reflection angles.

Scanning transmission electron microscopy (STEM) characterization

The metasurface samples for STEM characterization are fabricated with a substrate of highly boron-doped silicon wafer (University wafer, #1319) rather than glass slide for accessing overall high conductivity. To achieve cross-section imaging, we first made a lamella of the metasurface sample using a focused ion beam (FIB) system (ThermoFisher Helios 5CX (cryo) FIB-SEM). The lamella was prepared via the standard lift-out procedure, which includes using a gas injection system to coat the target area with a protective layer of platinum (Pt). Specifically, the Pt layer consisted of a ~300 nm electron-beam-deposited layer (5 kV, 1.4 nA) and a ~4.0 μm ion-beam-deposited layer (30 kV, 0.23 nA). An approximately 12 × 1.5 × 5.0 μm section was cut off from the patterned area using gallium (Ga) ion beams and attached to a Cu half grid using the EasyLift micromanipulator. The lamella was thinned on both sides (initially at 30 kV, 0.78 nA, then at 16 kV, 240 pA, and 5 kV, 47 pA) and then cleaned (at 2 kV, 24 pA) with Ga ion beams, which yielded a ~70 nm thick lamella. Afterwards, the STEM imaging of the lamella was done on the aberration-corrected JEOL ARM200CF with a cold field emission source operated at 200 kV. The STEM images were obtained with the high-angle annular dark-field detector (90-270 mrad). Both the FIB and STEM processes were done with the instruments at the University of Illinois at Chicago.

**Data availability**

All data needed to evaluate the conclusions in the paper are present in the paper and/or the Supplementary Materials.

**Acknowledgements**

This work made use of the shared facilities in the Searle Cleanroom and the Pritzker Nanofabrication Facility, part of the Pritzker School of Molecular Engineering at the University of Chicago, which is supported by Soft and Hybrid Nanotechnology Experimental (SHyNE) Resource (NSF-ECCS-2025633), a node of the National Science Foundation's National Nanotechnology Coordinated Infrastructure (RRID: SCR_022955). Parts of this work were carried out with the instruments in the Electron Microscopy Core, Research Resources Center at the University of Illinois at Chicago. Q.L. acknowledges Wook Jun Nam for the help in the nanofabrication process and Zhe Cheng for the help with the optical microscope characterization.

**Funding:** The project was supported by the startup fund from the Pritzker School of Molecular Engineering at the University of Chicago and the National Science Foundation (Electrical, Communications and Cyber Systems award no. 2145933). K.A. acknowledges support from the Air Force Office of Scientific Research under award number FA9550-22-1-0300.

**Author contributions**: P.-C.H. and Q.L. conceived the idea. Q.L. performed the device design and simulations, fabrications, electrochemical experiments, characterizations, and corresponding data analyses. S.P.K and K.A. conducted the angle-resolved spectroscopy measurements, C.S. assisted with the electrochemical experiments, T.-H.C. and X.W. helped with the ellipsometer measurements, G.Y. carried out the lamella preparation using FIB, STEM, and EDS characterizations, R.W. and P.-J.H. assisted with sample preparation, W.C. helped with the e-beam lithography, T.E. contributed to the Fourier plane imaging. Q.L. and P.-C.H. wrote the manuscript with input from all co-authors.

**Conflict of interest**

The authors declare they have no competing interest.

102. Z. Li, C. Wan, C. Dai, J. Zhang, G. Zheng, Z. Li, Actively Switchable Beam-Steering via Hydrophilic/Hydrophobic-Selective Design of Water-Immersed Metasurface. *Adv. Opt. Mater.* **9**, 2100297 (2021).

103. J. N. Fuller, Thomas F., Harb, Electrochemical Engineering. John Wiley & Sons Ltd (2018).

104. J. Kischkat, S. Peters, B. Gruska, M. Semtsiv, M. Chashnikova, M. Klinkmüller, O. Fedosenko, S. Machulik, A. Aleksandrova, G. Monastyrskyi, Y. Flores, W. Ted Masselink, Mid-infrared optical properties of thin films of aluminum oxide, titanium dioxide, silicon dioxide, aluminum nitride, and silicon nitride. *Appl. Opt.* **51**, 6789–6798 (2012).

105. S. Babar, J. H. Weaver, Optical constants of Cu, Ag, and Au revisited. *Appl. Opt.* **54**, 477–481 (2015).

106. A. G. Mathewson, H. P. Myers, Absolute Values of the Optical Constants of Some Pure Metals. *Phys. Scr.* **4**, 291 (1971).
35